\documentclass[12pt,a4paper]{article}
\usepackage[english]{babel}
\usepackage[latin1]{inputenc}
\usepackage[T1]{fontenc}
\usepackage{amsmath}
\usepackage{amsfonts}
\usepackage{amssymb}
\usepackage{braket}
\usepackage{graphicx,color}
\usepackage{hyperref}
\usepackage{tikz}
\usepackage{lmodern,bm}
\usepackage{pgfplots}
\usepackage{slashed}
\usepackage{dsfont}
\pgfplotsset{compat=1.17}

\begin{document}
\title{A functional approach to the Van der Waals interaction} 
\author{C.~D.~Fosco and G.~Hansen\\
{\normalsize\it Centro At\'omico Bariloche and Instituto Balseiro}\\
{\normalsize\it Comisi\'on Nacional de Energ\'{\i}a At\'omica}\\
{\normalsize\it R8402AGP S.\ C.\ de Bariloche, Argentina.}}

\maketitle
\begin{abstract} 
Based on a microscopic model, we use a functional integral approach to
evaluate the quantum interaction energy between two neutral atoms.  Each
atom is coupled to the electromagnetic (EM) field via a dipole term,
generated by an electron bound to the nucleus via a harmonic potential. We
show that the resulting expression for the energy becomes the Van der Waals
interaction energy at the first non-trivial order in an expansion in powers
of the fine structure constant, encompassing both the long and short distance 
behaviours. 
We also explore the opposite, strong-coupling limit, which yields a result for
the interaction energy as well as a threshold for the existence of a vacuum 
decay probability, manifested here as an imaginary part for the effective action.

In the weak-coupling limit, we also study the effect of using a general central
potential for the internal structure of the atoms.
\end{abstract}
\section{Introduction}\label{sec:intro}
A celebrated manifestation of the existence of vacuum fluctuations are the
Casimir, Van der Waals, and related interactions~\cite{milonni,milton,bordag}.
The second is a well-known example of an attractive force, between two neutral atoms, 
which results from the correlation between their dipole-moment fluctuations.  
That correlation, on the other hand, is mediated by the (vacuum) electromagnetic (EM) 
field.  

In this paper, we use a microscopic model to derive the interaction energy
between two neutral atoms, each one described by a static nucleus, to which
an electron is bound by a harmonic potential. The coupling of each atom to
the EM field is, on the other hand, implemented by a dipole term.
In the approach that we follow, we evaluate the interaction energy by
calculating the Euclidean (imaginary time) effective action resulting from
the integration of the quantum fluctuations of the electrons and the EM
field.  The vacuum energy obtained thusly, may be thought of as the
result of taking the zero-temperature limit of the thermal free energy.

This paper is organized as follows: in Section~\ref{sec:themodel} we
introduce the model we use to describe the system, and the tools used to
evaluate the interaction energy, in particular, its imaginary-time effective
action. Then, in Sect.~\ref{sec:ei}, we evaluate the static interaction 
energy between the two atoms, discussing different limits. We also derive an
expression for the imaginary part of the energy, and interpret it in terms
of a vacuum decay probability. We conclude the section by studying the
case of a general central potential for the atoms, in the weak-coupling regime.

Finally, in Sect.~\ref{sec:conc}, we present our conclusions.

\section{The system and its effective action}\label{sec:themodel}
\subsection{The model}
The model that we consider in this work deals with two atoms, labeled by
$1$ and $2$, having their centres of mass at ${\mathbf r}^{(1)}$ and
${\mathbf r}^{(2)}$, while the electrons are located at the positions
${\mathbf x}^{(1)}$ and ${\mathbf x}^{(2)}$, relative to ${\mathbf
r}^{(1)}$ and ${\mathbf r}^{(2)}$, respectively.  The action ${\mathcal
S}$,  a functional of the gauge field $A$ also, is given by: 
\begin{align}
{\mathcal S}({\mathbf x}^{(1)},{\mathbf x}^{(2)}, \,A\;;
{\mathbf r}^{(1)},{\mathbf r}^{(2)})
&=\; 
{\mathcal S}^a_0({\mathbf x}^{(1)}) + {\mathcal S}^a_0({\mathbf x}^{(2)})  \,+\,
{\mathcal S}^a_I({\mathbf x}^{(1)}, \, A\; ;{\mathbf r}^{(1)} ) \nonumber\\
&+\, {\mathcal S}^a_I({\mathbf x}^{(2)}, \, A\; ; {\mathbf r}^{(2)} )
\,+\, 
{\mathcal S}^{\rm EM}_0(A)\;,
\end{align}
where ${\mathcal S}^a_0({\mathbf x})$ is the action for an electron in the
presence of the bounding potential, 
${\mathcal S}^{\rm EM}_0(A)$ the one for the free EM field, and ${\mathcal
S}^a_I$ contains the coupling of an electron to the EM field. The ${\mathbf r}^{(1)}$ 
and ${\mathbf r}^{(2)}$ vectors, which appear in the action, are to be 
regarded as external parameters: the ones upon which the effective
action will depend.

For the sake of simplicity, the action for each orbiting electron, having mass $m$
and position $\mathbf{x}(t)$ relative to the nucleus, is taken to be of the form:
\begin{equation}\label{eq:defsah}
{\mathcal S}^{a}_0({\mathbf x})\;=\; \frac{m}{2} \,  \int dt \,
\big( \dot{\mathbf{x}}^2 - \Omega^2 {\mathbf x}^2 \big) \;,
\end{equation}
since it will allow for the exact evaluation of the interaction energy. Note, 
however, that it may be applied to some real physical systems, like heavy muonic 
atoms~\cite{muonic1,muonic2}. 
 
The interaction with the EM field is assumed to be given by
the dipolar term:
\begin{equation} \label{eq:SaI}
    \mathcal{S}_{I}^{a}(A; \mathbf{r}, \mathbf{x})
    = \int dt \, q \, x_i \, E_i(\mathbf{r}) \;,
\end{equation}    
where $q$ is the charge of the ``electron''~\footnote{We use this terminology,
although the actual value of $q$ will be assumed to be a variable which measures
the strength of the EM coupling. In the same vein, the binding potential is 
not Coulombian but harmonic}, and $E_i$ denotes the $i^{th}$
component of the electric field. In our conventions, $E_i = F_{0i}$ with
\mbox{$F_{\mu\nu} = \partial_\mu A_\nu - \partial_\nu A_\mu$}.  Indices
from the middle of the Latin alphabet: $i, \, j, \, k, \ldots$ run from $1$
to $3$, while Greek ones are assumed to run from $0$ to $3$.  Besides, we
shall later on use $\alpha, \beta, \ldots$, taking values $1$ and $2$, corresponding
to the two atoms.

We follow Einstein's convention: throughout this paper, a sum over repeated
indices is assumed unless explicitly stated otherwise.

Here, $A_\mu$ denotes the 4-potential, which has the action:
\begin{equation}
\mathcal{S}_{0}^{\text{EM}}(A) = \int d^4 x \,
\left[-\frac{1}{4} F_{\mu\nu} F^{\mu\nu} - \frac{\lambda}{2} \,
(\partial_{\mu} A^{\mu})^2\right] \;\;,
\end{equation}
consisting of the standard (vacuum) Maxwell term, plus a covariant
gauge-fixing term ($\lambda \neq 0$).  We use natural units ($c = 1$,
$\hbar = 1$) and the metric signature $(+, -, -, -)$.

\subsection{Effective action}\label{sec:effective_act}
 A rather convenient way to obtain the quantum interaction energy for a
 system composed of two or more objects is by means of its imaginary-time
 effective action, $\Gamma_{\text{eff}}$. Indeed, by considering a static
 configuration, one can extract the vacuum energy by taking the limit:
\begin{equation}
	E_I \;=\; \lim_{T \to \infty} 
	\left(\frac{\Gamma_{\text{eff}}}{T} \right) \;,
\end{equation}
where $T$ denotes the extent of the (imaginary) time interval \cite{ZinnJustin}.
$\Gamma_{\text{eff}}$ 
results from the integration of the quantum fluctuations, yielding as a result a 
function of the remaining, classical degrees of freedom. Since we are
interested in the {\em interaction\/} part of the energy, we shall subtract
the self-energy contributions, which are the ones that survive when the
objects are infinitely far apart.

We will use for $\Gamma_{\text{eff}}$ a convenient representation in terms of a
functional integral:
\begin{align}\label{eq:defgammaeff_0}
& e^{-\Gamma_{\text{eff}}(\mathbf{r}^{(1)}, \mathbf{r}^{(2)})} \,\equiv \,  
\frac{1}{\mathcal N} 
\; {\mathcal Z}(\mathbf{r}^{(1)}, \mathbf{r}^{(2)}) \nonumber\\
& {\mathcal Z}(\mathbf{r}^{(1)}, \mathbf{r}^{(2)}) =\,\int \mathcal{D}\mathbf{x}^{(1)} \, 
\mathcal{D}\mathbf{x}^{(2)} \, \mathcal{D}A \, 
\,
e^{-\mathcal{S}_E(\mathbf{x}^{(1)},\,\mathbf{x}^{(2)},\,A\,;\,\mathbf{r}^{(1)},\,
\mathbf{r}^{(2)})} \;,
\end{align}
where ${\mathcal N}$ is a constant, and $\mathcal{S}_E$ is the Euclidean
(Wick rotated) version of the action:
\begin{align}
\mathcal{S}_E \,=\, \frac{m}{2} \, \int d\tau \, \Big[ ( \dot{\mathbf x}^{(1)} )^2 
+ ( \dot{\mathbf x}^{(2)} )^2 &+\, 
	\Omega^2 \big( ({\mathbf{x}}^{(1)})^2
+ ({\mathbf{x}}^{(2)})^2  \big) \Big] \nonumber\\
	\,+\,q \, \int d\tau \, \big( {\mathbf x}^{(1)} \cdot \, {\mathbf E}(\tau,\mathbf{r}^{(1)})
\,+\, {\mathbf x}^{(2)} \cdot \, {\mathbf E}(\tau,\mathbf{r}^{(2)})\big)
\,
& + \,  \int d^4 x \, \frac{1}{2} A_\mu (-\partial^2) A_\mu \;.
\end{align}    
Here, $\tau \equiv x_0$ is the imaginary time, the metric becomes
\mbox{$(g_{\mu\nu}) = {\rm diag}(1,1,1,1)$}, ${\mathbf E} \equiv \partial_\tau
{\mathbf A} - \nabla A_0$, and we have adopted the Feynman ($\lambda = 1$) gauge.
The normalization constant ${\mathcal N}$, is chosen
in such a way that the energy vanishes when the distance between atoms
tends to infinity:
\begin{equation}
{\mathcal N} \,=\, \left[{\mathcal Z}(\mathbf{r}^{(1)}, \mathbf{r}^{(2)})
\right]\big|_{|\mathbf{r}^{(1)} - \mathbf{r}^{(2)}| \to \infty} \;.
\end{equation}
Note that this implies that any factor independent of $\mathbf{r}^{(1)}$
or $\mathbf{r}^{(2)}$ in ${\mathcal Z}$, may be discarded.

As a first step towards obtaining $\Gamma_{\text{eff}}$, we introduce an 
intermediate object, ${\mathcal S}_{\text{eff}}$: the result of performing
the functional integral just over $A_\mu$:
\begin{align}\label{eq:seff_1}
	e^{-{\mathcal S}_{\text{eff}}(\mathbf{x}^{(1)},\, \mathbf{x}^{(2)}\,;\,
		\mathbf{r}^{(1)},\,
    \mathbf{r}^{(2)})} 
& =\, 
e^{-\frac{m}{2} \int_\tau \,\big[ ( \dot{\mathbf x}^{(1)} )^2 + ( \dot{\mathbf x}^{(2)} )^2 \,+\, 
\Omega^2 \big( ({\mathbf{x}}^{(1)})^2
+ ({\mathbf{x}}^{(2)})^2  \big)\big]} \nonumber\\
& \times \, e^{-\frac{1}{2} \int_{x,y} J_\mu(x) \Delta_{\mu\nu}(x-y)J_\nu(y)} \;,
\end{align}
where we used a shorthand notation for the integrations, and $J_\mu =
J_\mu^{(1)} + J_\mu^{(2)}$, $J_\mu^{(\alpha)}$ ($\alpha =1,\,2$) being
a dipole current concentrated on each atom, given explicitly by:
\begin{align}\label{eq:defcurrent}
J_0^{(\alpha)}(y) & =\, - q \, x^{(\alpha)}_j(\tau) \, \delta(y_0 - \tau) \, 
\frac{\partial}{\partial y_j} \,\delta^3({\mathbf y}- {\mathbf r}^{(\alpha)}) \nonumber\\
{\mathbf J}^{(\alpha)}(y) & =\,  q \, \dot{\mathbf x}^{(\alpha)}(\tau) \, \delta(y_0-\tau) \,
\delta^3({\mathbf y} - {\mathbf r}^{(\alpha)}) \;\;,
\end{align}
and
\begin{equation}
	\Delta_{\mu\nu}(x - y) =  \delta_{\mu\nu} \, \Delta(x-y) \,
\end{equation}
where $\Delta(x - y)$ is the scalar propagator:
\begin{equation}\label{eq:scalarprop}
\Delta(x - y) = \int \frac{d^4 k}{(2\pi)^4} \,
 \frac{e^{- i k (x-y)}}{k^2} \;.
\end{equation}
Note that each current is conserved ($\partial_\mu J_\mu^{(\alpha)} = 0$), so
the result (\ref{eq:seff_1}) is independent of the value of the constant
$\lambda$. Taking into account the form of the gauge field propagator one
sees that, due to the coupling to the EM field, $\Omega$ in the harmonic
term of each atom's action gets renormalized.  Keeping the
same notation, $\Omega$, now for the {\em renormalized\/} frequency, we see that 
\begin{equation}\label{eq:gammaeff_1}
{\mathcal S}_{\rm eff}({\mathbf x}^{(1)}, {\mathbf x}^{(2)}, A \,; {\mathbf r}^{(1)},
{\mathbf r}^{(2)} ) \;=\;\frac{1}{2} \, \int_{-\infty}^{+\infty} \frac{d\nu}{2\pi}\, 
[\tilde{x}^{(\alpha)}_i(\nu)]^* \,K_{ij}^{(\alpha \beta)}(\nu)\,
\tilde{x}^{(\beta)}_j(\nu) \;,
\end{equation}
where we have introduced: 
\begin{equation}
\tilde{x}^{(\alpha)}_j(\nu) \equiv \int_{-\infty}^{+\infty} dt \, e^{i \nu
t} \, x^{(\alpha)}_j(t) \;\;,
\end{equation}
and $^*$ denotes complex conjugation.

On the other hand,
\begin{equation}
	K_{ij}^{(ab)}(\nu) \; \equiv \; m  (\nu^2 \,+\, \Omega^2)
	\delta^{(ab)} \delta_{ij} \;+\; \sigma^{(ab)}  \, M_{ij}(\nu)
\end{equation}
where $\sigma^{(ab)} \,\equiv\,  \delta^{(a 1)} \delta^{(b 2)}
+ \delta^{(a 2)} \delta^{(b 1)}$, while the matrix elements $M_{ij}(\nu)$
are given by:
\begin{equation}
 M_{ij}(\nu) \equiv q^2 \int \frac{d^3\mathbf{k}}{(2\pi)^3}
 \, e^{i \mathbf{k} \cdot \mathbf{r}} \, 
 \frac{\nu^2 \delta_{ij} + k_i k_j}{\nu^2 + \mathbf{k}^2} \;\;,\;\;\;\;
	\mathbf{r} \equiv \mathbf{r}^{(1)} - \mathbf{r}^{(2)} \;.
\end{equation}
With the appropriate choice of the system of coordinates ($\hat{e}_{k_3}
\, || \, \mathbf{r}$), we cast this Hermitian matrix in diagonal form:
${\rm diag}(M_1(\nu), M_2(\nu), M_3(\nu))$, with:
\begin{align}
M_1(\nu) & = M_2(\nu) = - \frac{q^2}{4 \pi r^3} \, (1 + r |\nu|
+ r^2 |\nu|^2) \, e^{- r |\nu|}, \nonumber\\
M_3(\nu) & = - \frac{q^2}{2 \pi r^3} \, (1 +  r |\nu|) \,
 e^{- r |\nu|} \;,
\end{align}
where, as expected, the dependence on the relative positions of the atoms 
is only through their distance: $r \equiv |{\mathbf r}|$.

Finally, we integrate out the electrons' coordinates relative to each atom. This 
is still a Gaussian functional integral which, by converting to Fourier space 
also the integration measure, becomes an infinite product of decoupled 
ordinary integrals. This means that the integral
\begin{equation}\label{eq:defgammaeff}
 e^{-\Gamma_{\text{eff}}(r)} \,=\,  
\frac{1}{\mathcal N} \int \mathcal{D}\mathbf{x}^{(1)} \,
\mathcal{D}\mathbf{x}^{(2)}\, e^{- {\mathcal S}_{\text{eff}}({\mathbf x}^{(1)}, 
{\mathbf x}^{(2)}, A \,; r )} \;,
\end{equation}
yields an expression for $\Gamma_{\text{eff}}$ which is proportional to the
total evolution time, $T$, as it should be for a static configuration. The
interaction energy, however, is given by the ratio
$\frac{\Gamma_{\text{eff}}}{T}$ in the $T \to \infty$ limit, which is
well-defined: 
\begin{equation}
E_I(r) \;=\; \Big[\frac{{\Gamma}_{\text{eff}}({\mathbf r})}{T}\Big]_{T \to
\infty} \, = \,  
\frac{1}{2} \, \int_{-\infty}^{\infty}\frac{d\nu}{2\pi} \, \log \det (\mathds{1} - \mathds{T}(\nu)),
\end{equation}
where $\mathds{T}(\nu)$ is the $3\times 3$ matrix:
\begin{equation}
	\mathds{T}(\nu) \, = \, e^{- 2 |\nu| r} \, 
    \begin{bmatrix}
	    [\xi_\shortparallel (\nu, r)]^2  & 0 & 0 \\
	    0 & [\xi_\shortparallel (\nu, r)]^2 & 0 \\
	    0 & 0 & [\xi_\perp (\nu, r)]^2  
    \end{bmatrix}\; ,
\end{equation}
where 
\begin{equation}
\xi_\shortparallel (\nu, r) \,=\, \frac{q^2 (1 + r |\nu|
+ r^2 |\nu|^2)}{4 \pi r^3 m (\nu^2 + \Omega^2)} \;\;,\;\;\;
\xi_\perp (\nu, r) \,=\, \frac{q^2 (1 + r |\nu|)}{2 \pi r^3 m (\nu^2 +
	\Omega^2)} \;. 
\end{equation}
\section{Interaction energy}\label{sec:ei}
Computing the determinant, we have the resulting expression for the 
interaction energy
\begin{equation}\label{eq:interaction}
E_I(r) = \frac{1}{2}
\int_{-\infty}^{\infty} \frac{d\nu}{2\pi} \,
\big[2 \log\left( 1 - \xi_\shortparallel^2 e^{- 2 |\nu| r}  \right)
+ \, \log\left( 1 - \xi_\perp^2 e^{- 2 |\nu| r} \right) \big]\;
\end{equation}
which is a realization, in a particular context, of the $TGTG$ formula~\cite{klich}.

By a redefinition of the integration variable: $\nu = \frac{u}{r}$, we see
that:
\begin{align}
E_I(r) = \frac{1}{2 r}
\int_{-\infty}^{\infty} \frac{du}{2\pi} \,
\Big\{2 &\log\big[ 1 - \big(\frac{q^2}{4 \pi  m \, r} 
\frac{1 + |u| + u^2}{u^2 + (\Omega r)^2}\big)^2  e^{- 2 |u|} \big] \nonumber\\
+ \,& \log\big[ 1 - \big( \frac{q^2}{2 \pi  m \, r} \frac{1 + |u|}{u^2 +
(\Omega r)^2}\big)^2 e^{- 2 |u|}  \big]  \Big\} \;.
\end{align}
Introducing the dimensionless variable $x \equiv \Omega r$, which measures
the distance between atoms in terms of a length scale $\sim \Omega^{-1}$, and 
using $\Omega$ to measure energies: $E_I(r) \equiv \Omega {\mathcal E}_I(\Omega r)$,
where:
\begin{align}
{\mathcal E}_I(x) = \frac{1}{x}
\int_0^{\infty} \frac{du}{2\pi} \,
\Big\{2 &\log\big[ 1 - \big(\frac{q^2}{4\pi} \frac{\Omega}{m}\big)^2 \,  
 \frac{(1 + u + u^2)^2}{x^2 ( u^2 + x^2)^2}  \, e^{- 2 u} \big] \nonumber\\
 + \,& \log\big[ 1 - \big(\frac{q^2}{2\pi} \frac{\Omega}{m}\big)^2 \, 
 \frac{(1 + u)^2}{x^2 (u^2 + x^2)^2} \,  e^{- 2 u} 
 \big]  \Big\} \;,
\end{align}    
a result that we shall analyze below under different assumptions
regarding the parameters of the system.

\subsection{Weak coupling and Van der Waals interaction} 
This corresponds to situations where, keeping a few terms in the power series
expansion: \mbox{${\rm log}(1 - x) = - \sum_{n=1} \frac{x^n}{n+1}$} 
for each one of the logs above, is a reliable approximation.
  
Moreover, we also assume here that that is achieved
by means of the two (independent) conditions: 
$\frac{q^2}{2\pi} \ll m/\Omega$, and $r \gg 1/\Omega$. 
The first one is essentially a constraint on the maximum value of the
coupling constant or, equivalently, on the size of the electric dipole 
fluctuations on each atom. 

\begin{figure}
\begin{center}
\includegraphics[width=0.67\textwidth]{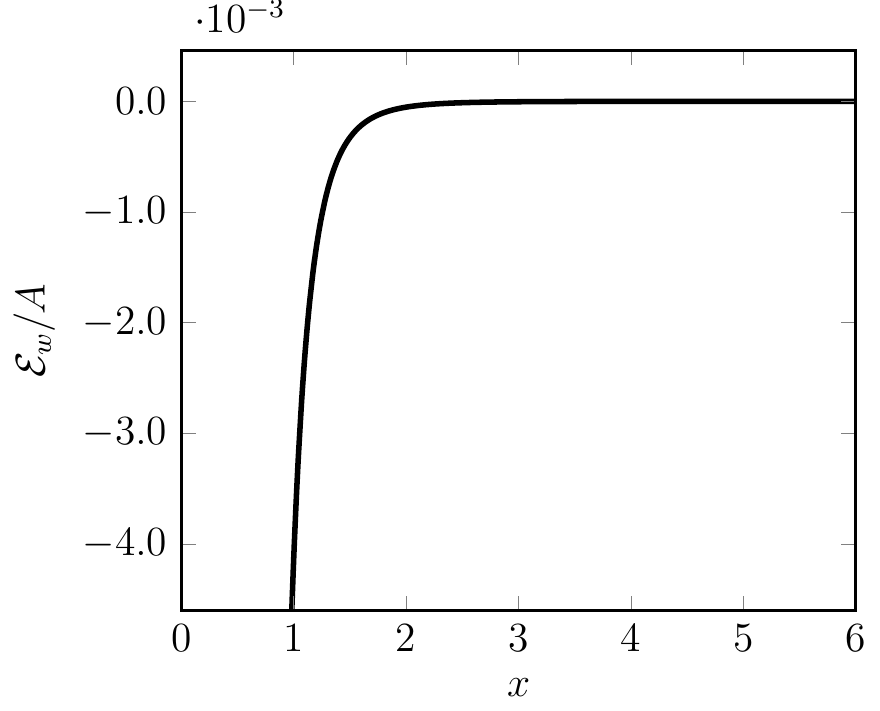}
\caption{Weak coupling interaction energy ($\mathcal{E}_{w}$) as function
of $x = \Omega r$. $A$ is the global factor $A = \big(\frac{q^2 \Omega}{m}\big)^2$.}
\label{fig:Ew}
\end{center}         
\end{figure}

Under those two assumptions, the leading term in the expansion is:
\begin{align}
\label{eq:EwInt}
{\mathcal E}_I(x) & \sim  {\mathcal E}_w(x)  \nonumber\\
{\mathcal E}_w(x) &\equiv \, - \frac{1}{16 \pi^3} 
\big(\frac{q^2 \Omega}{m}\big)^2  \, \frac{1}{x^3}\, 
\int_0^{\infty} du \,e^{- 2 u} \, \frac{u^4 + 2 u^3 + 5 u^2 + 6 u + 3}{(u^2
+ x^2)^2} \;.
\end{align}
The integral from (\ref{eq:EwInt}) can be computed exactly, for that we introduce
the two auxiliary functions:
\begin{align}
f(x) &\equiv \text{Ci}(x) \sin(x) - \text{si}(x) \cos(x), \\
g(x) &\equiv -(\text{Ci}(x) \cos(x) + \text{si}(x) \sin(x)),
\end{align}
where $\text{si}(x) \equiv \text{Si}(x) - \frac{\pi}{2}$, being $\text{Ci}(x)$ and
$\text{Si}(x)$ the cosine and sine integrals, respectively \cite{AbramowitzStegun}.
In terms of those functions, we have:
\begin{align}
\label{eq:Ew}
\mathcal{E}_w(x) = -\frac{1}{32\pi^3} \left(\frac{q^2 \Omega}{m}\right)^2 \frac{1}{x^6} \,
&\big(x\,(6 - x^2) + (3 - 7x^2 + x^4) \, f(2x) \nonumber\\
&+ 2x\,(3 - 3x^2 + x^4) \, g(2x) \big).
\end{align}
We plot $\mathcal{E}_w(x)$ in Fig. \ref{fig:Ew}.
For $x \gg 1$, result (\ref{eq:Ew}) reproduces the asymptotic behaviour of
Van der Waals forces at long distances:
\begin{equation}
\label{eq:EIVDW}
E_{I}(r) \sim -\frac{23}{(4\pi)^3} \,
\big(\frac{q^2}{m \Omega^2}\big)^2 \, \frac{1}{r^7} \;,
\end{equation}
in agreement with \cite{Feinberg} (see also \cite{ItzyksonZuber}). From (\ref{eq:EIVDW}),
we identify the static electric susceptibility of the microscopic model:
$\alpha_E = \frac{q^2}{m \Omega^2}$, which has volume dimensions.

In terms of the original variables, and written in a way that makes the comparison
with the next-to-leading term more straightforward,
\begin{equation}
E_I(r) \sim -\frac{23}{4\pi} \,
\big(\frac{q^2}{4 \pi m \Omega}\big)^2 \,\frac{1}{\Omega r} \;
\frac{1}{\Omega r^6} \;.
\end{equation}

The next-to-leading term at long distances corresponds to the London limit~\cite{London,milonni}. This
can be picked up by extracting the next negative power of the distance or,
equivalently, by evaluating the frequency integral in (\ref{eq:interaction}) 
using the approximation $|\nu| r \simeq 0$. In this situation,
\begin{align}
E_I(r) & \simeq\, -\frac{1}{2}\, 
\int_{-\infty}^{+\infty} \frac{d\nu}{2\pi} \, \frac{1}{( \nu^2 + \Omega^2)^2}
\, 
\Big[
2 \, \big(\frac{q^2}{4 \pi r^3 m} \big)^2 + \big(\frac{q^2}{2\pi r^3 m} \big)^2
\Big] 
\nonumber\\
& = \; - \frac{3}{4} \, \big(\frac{q^2}{4 \pi m \Omega} \big)^2 \,
\frac{1}{\Omega \, r^6}
\end{align} 

The ratio between this term and the asymptotic one at long distances is 
${\mathcal O}(\Omega r)$, as it should be.

\subsection{Strong coupling: short distances and imaginary part of the energy} 
One should expect the interaction energy to be real. Note, however, that due to 
the presence of the logarithms, the interaction will have both real 
and imaginary parts. The existence of imaginary parts for the logarithms 
may be seen to depend on the value of the dimensionless ratio:
\begin{equation}
g\; \equiv \;
\frac{q^2}{2\pi} \frac{1}{m \Omega^2 r^3} \;.
\end{equation}

The existence of an instability at short distances, and therefore of an 
imaginary part in the effective action, may be understood by an argument based on the
form of the intermediate effective action ${\mathcal S}_{\text{eff}}$, 
and its short distance behaviour: the London limit. 
We recall that it is a quadratic form in the electrons' coordinates, with 
a frequency-dependent kernel $K_{ij}^{(ab)}(\nu)$.
At sufficiently small distances $r$, we may use the instantaneous approximation
for the gauge field propagator, namely, assume that $r \ll \lambda$, and therefore 
use $r \nu \simeq 0$ (see \cite{milonni} \& references there in). 

This is in fact equivalent to replacing the gauge-field propagator by its Coulomb
 form. Therefore,
\begin{equation}
	K_{ij}^{(ab)}(\nu) \; \simeq\;  m  (\nu^2 \,+\, \Omega^2)
	\delta^{(ab)} \delta_{ij} \;+\; \sigma^{(ab)}  \, M_{ij}(0)
\end{equation}
where $M_{ij}(0)$ is diagonal if the orthogonal coordinate system is chosen so that $x_3$ 
points along the direction of the line connecting the two atoms. With this choice,
\begin{equation}
 \big[ M_{ij}(0) \big] \,\equiv\, 
 - \frac{q^2}{4 \pi r^3} 
 \,
 \left( \begin{array}{ccc}
 1 & 0 & 0 \\
 0 & 1 & 0 \\
 0 & 0 & 2
 \end{array}
\right)
\;.
\end{equation}
In this limit, transforming back the electrons fluctuations from frequency to time, 
${\mathcal S}_{\text{eff}}$ becomes local in time, and may thus be interpreted as 
an action. 

That action involves the original six harmonic modes (one for each
$x^{(\alpha)}_i$) with 
identical oscillation frequency $\Omega$, plus an $r$-dependent term which couples 
them. Altogether, they produce a potential which we denote by
$V_{\text{eff}}$, and is still quadratic:   
\begin{align}
	\label{eq:Seff}
	\mathcal{S}_{\text{eff}} & \simeq\, \int d\tau \, \big[ \frac{m}{2} 
	\dot{x}^{(\alpha)}_i \dot{x}^{(\alpha)}_i \,+\,
	V_{\text{eff}}( \{ x^{(\alpha)}_i \} )
\big] \nonumber\\
V_{\text{eff}}(\{ x^{(\alpha)}_i \}) & =\,\frac{m}{2} \Omega^2
x^{(\alpha)}_i x^{(\alpha)}_i 
- \frac{q^2}{4 \pi r^3} \big( x^{(1)}_1 x^{(2)}_1 + x^{(1)}_2 x^{(2)}_2 
+ 2 x^{(1)}_3 x^{(2)}_3 \big) \;.
\end{align}
The diagonalization of this  potential is straightforward; the normal coordinates
\begin{align}\label{eq:normal}
x_i^{(\pm)} \;\equiv\; \frac{x^{(1)}_i \pm  x^{(2)}_i}{\sqrt{2}}  \;\;,\;\;\; 
i = 1,\, 2,\,3 \;,
\end{align}
lead to the potential:
\begin{align}
	\label{eq:Veff}
	V_{\text{eff}}(\{ x^{(\pm)}_i \}) &=\,\frac{m}{2} \sum_{i=1}^3 \, \big( (\Omega^{+}_i)^2 \, x^{(+)}_i x^{(+)}_i
	+ (\Omega^{-}_i)^2 \, x^{(-)}_i x^{(-)}_i \big) \nonumber\\
(\Omega^\pm_1)^2  &=\;  (\Omega^\pm_2)^2 \,=\, \Omega^2 \,\mp\, \frac{q^2}{4 \pi m r^3} \;\;,
\;\;\; (\Omega^\pm_3)^2 \,=\,  \Omega^2 \,\mp\, \frac{q^2}{2 \pi m r^3}  \;.
\end{align}

By diagonalizing the action (\ref{eq:Seff}), the system becomes a set of three
uncoupled harmonic oscillators, which partition function factorices as the product
of the partition functions of each individual oscillator, with frequencies
$\sqrt{(\Omega_{i}^\pm)^2}$, $i = 1, 2, 3$ \cite{ZinnJustin,Schulman}.
The energy of the system is:
\begin{equation}
	E_0 = \frac{1}{2}\,\sum_{i = 1}^{3} \big( \sqrt{{(\Omega_{i}^{+})}^2}
	+ \sqrt{{(\Omega_{i}^{-})}^2} \, \big).
\end{equation}

\begin{figure}
\begin{center}
\includegraphics[width=0.67\textwidth]{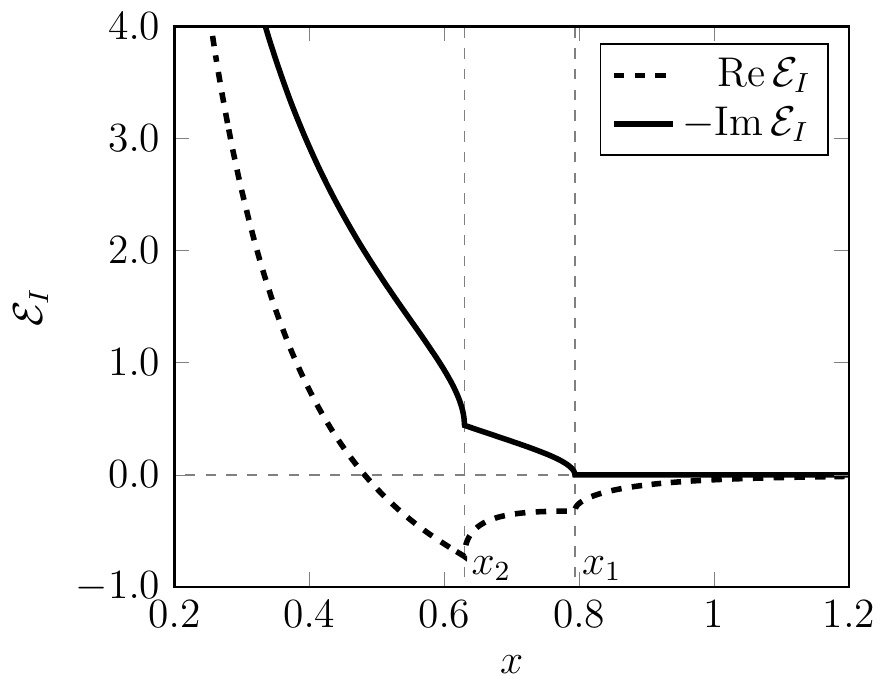}
\caption{Real and imaginary parts of the interaction energy ($\mathcal{E}_{I}$), as
a function of $x = \Omega r$, with $\frac{q^2}{2\pi} \, \frac{\Omega}{m} = 0.5$.
$x_1 = \Omega r_1$ and $x_2 = \Omega r_2$ are the first and second thresholds
for which $\mathcal{E}_{I}$ develops an imaginary part.}
\label{fig:EI}
\end{center}
\end{figure}

From the expressions of (\ref{eq:Veff}), we see that there will appear complex 
frequencies, depending on the value of $r$. Defining the two distances: 
\begin{equation}
r_1 \, \equiv \, \big( \frac{q^2}{2\pi m \Omega^2} \big)^{1/3} \;\;,\;\;\;
r_2 \, \equiv \, \big( \frac{q^2}{4\pi m \Omega^2} \big)^{1/3} \;\;,
\end{equation}
we note that there is a first threshold at $r = r_1$ for the existence of an 
complex frequency, and then another one at $r = r_2 < r_1$. Note that this explains
the behaviour observed in the plot of the imaginary part of the energy
as a function of $r$. Indeed, besides the clear existence of the first threshold at
$r = r_1$, we also see the emergence of the second one at $r = r_2$. Note also that
the existence of two modes for the latter is reflected in the steepest rise
for $r < r_2$. The real and imaginary parts of the interaction energy are shown in
Fig. \ref{fig:EI}.

The physical interpretation of the imaginary part of the energy is that of a
non-vanishing probability of vacuum decay, understanding as vacuum the one used in 
the calculation of the effective action. Regarding the atoms, that vacuum is the 
tensor product of the two respective ground states. 
On the other hand, when the two atoms are sufficiently close to each other, 
it is clear the true vacuum should be closer to the one of two electrons in a
molecule. That is not a tensor product of the two: rather, it should be closer
to a linear combination of two atomic orbitals. That is indeed what may be seen 
from the form of the normal coordinates obtained in (\ref{eq:normal}): the modes
that destabilize the vacuum  correspond to 
${\mathbf x}^{(+)} = \frac{x^{(1)} +  x^{(2)}}{\sqrt{2}}$.

\subsection{General central potential}\label{sec:general}
Let us consider here the case of a more general central potential $V$, in such a
way that the action for each atom, rather than having the specific form
(\ref{eq:defsah}) is now assumed to fall under the more general form:
\begin{equation}\label{eq:gensa}
{\mathcal S}^{a}_0({\mathbf x})\;=\; \,  \int dt \,
\big[ \frac{m}{2} \, \dot{\mathbf{x}}^2 - V(|{\mathbf x}|) \big] \;.
\end{equation}
The most immediate way to compute the effect of using this potential rather
than the original, harmonic one, is to evaluate the Euclidean the effective
action for the redefined action in the weak-coupling regime. The lowest
non-trivial contribution to $\Gamma_{\text{eff}}(r)$ is again of order
$q^4$,
\begin{equation}
\Gamma_{\text{eff}}(r) \;\simeq\;  \Gamma^{(4)}_{\text{eff}}(r) \;,  
\end{equation}
and it may be obtained by using the properly redefined ${\mathcal S}_{\text{eff}}$ 
in (\ref{eq:defgammaeff}), after integrating out the EM field fluctuations,
and discarding self-energy terms, the result being
\begin{align}\label{eq:gamma4}
\Gamma^{(4)}_{\text{eff}}(r) \;=\; -\frac{1}{2} \, \int d^4x \int d^4y & \int d^4z \int d^4w \, \Big[
\Delta(x-y) \Delta(z-w) \nonumber\\
 &\times \, \langle J_\mu^{(1)}(x) J_\nu^{(1)}(z) \rangle \,  
 \langle J_\mu^{(2)}(y) J_\nu^{(2)}(w) \rangle 
 \Big] 
 \;.
\end{align}
Here, the functional averaging is understood with the (\ref{eq:gensa}) action
determining the respective weight; namely: 
\begin{equation}
\langle J_\mu^{(\alpha)}(x) J_\nu^{(\alpha)}(y) \rangle \,\equiv\, 
\frac{\int {\mathcal D}{\mathbf x}^{(\alpha)} \; J_\mu^{(\alpha)}(x) J_\nu^{(\alpha)}(y) \,
e^{-{\mathcal S}_0^a({\mathbf x}^{(\alpha)})}}{\int {\mathcal D}{\mathbf
x}^{(\alpha)} e^{-{\mathcal S}_0^a({\mathbf x}^{(\alpha)})}}
\end{equation}
(no sum over $\alpha$). The model we are using is such that each electron
is concentrated on one of the atoms, and as a consequence  
\mbox{$\langle J_\mu^{(1)} J_\nu^{(2)}\rangle = 0$}.

Recalling the definition of the currents in (\ref{eq:defcurrent}), it is
clear that the above averages are going to depend on the correlators
involving the coordinates and velocities of the electrons:
\begin{align}
\langle x_i^{(\alpha)}(\tau) x_j^{(\alpha)}(\tau') \rangle \;,\;\;  
&\langle \dot{x}_i^{(\alpha)}(\tau)  x_j^{(\alpha)}(\tau') \rangle
\;\;, \nonumber\\
\langle \dot{x}_i^{(\alpha)}(\tau)  \dot{x}_j^{(\alpha)}(\tau') \rangle
\;,\;\;  &\langle \dot{x}_i^{(\alpha)}(\tau) x_j^{(\alpha)}(\tau') \rangle
\end{align}
(no sum over $\alpha$). 

Using all the ingredients above, a lengthy but otherwise straightforward
calculation allows us to evaluate (\ref{eq:gamma4}). We find that
it is proportional to the total time and, besides, it may be written as a
single integral over a frequency:
\begin{align}\label{eq:gamma4_1}
& \Big[\frac{\Gamma^{(4)}_{\text{eff}}(r)}{T}\Big]_{T \to \infty}
\;=\; -\frac{q^4}{2} \, \int \frac{d\nu}{2\pi} \, 
\widetilde{G}_{ij}(\nu) \, \widetilde{G}_{kl}(-\nu)
\Big[ 
	\partial_i \partial_k\widetilde{\Delta}(\nu,r)
	\, \partial_j \partial_l \widetilde{\Delta}(\nu,r) 
\nonumber\\
& - \nu^2 \, \delta_{jl} \,
\partial_i \partial_k \widetilde{\Delta}(\nu,r) 
\widetilde{\Delta}(\nu,r)
- \nu^2 \, \delta_{ik} \, \widetilde{\Delta}(\nu,r) \,
\partial_j \partial_l \widetilde{\Delta}(\nu,r) \nonumber\\
& + \nu^4 \, \delta_{ik} \delta_{jl} \, 
\widetilde{\Delta}(\nu,r) \widetilde{\Delta}(\nu,r)
\Big] \;,
\end{align}
where  $\widetilde{G}_{ij}(\nu)$ is the Fourier transform of 
$\langle x_i^{(\alpha)}(\tau) x_j^{(\alpha)}(\tau') \rangle$:  
\begin{equation}
\langle x_i^{(\alpha)}(\tau) x_j^{(\alpha)}(\tau') \rangle \,=\, 
\int \frac{d\nu}{2\pi} \, e^{i \nu (\tau - \tau')} \, \widetilde{G}_{ij}(\nu)
\;,
\end{equation}
and
\begin{equation}
	\widetilde{\Delta}(\nu,r) \;=\; \frac{e^{-|\nu| r}}{4 \pi r} \;.
\end{equation}

Under the assumption that the potential is central, we may, for each $\alpha$,
write:
\begin{align}
& \langle x_i^{(\alpha)}(\tau) x_j^{(\alpha)}(\tau') \rangle \,=\, \delta_{ij}
G(\tau - \tau') \;,\;\; 
\langle \dot{x}_i^{(\alpha)}(\tau) x_j^{(\alpha)}(\tau') \rangle \,=\,
\delta_{ij} \partial_\tau G(\tau - \tau') \nonumber\\
& \langle x_i^{(\alpha)}(\tau) \dot{x}_j^{(\alpha)}(\tau') \rangle \,=\, - \delta_{ij} 
\partial_\tau G(\tau - \tau')\;,\; 
\langle \dot{x}_i^{(\alpha)}(\tau) \dot{x}_j^{(\alpha)}(\tau') \rangle
\,=\, - \delta_{ij} \partial^2_\tau G(\tau - \tau') \;, 
\end{align}
in terms of a single scalar function $G$ (we recall that the atoms are
assumed to be identical).

Therefore, the energy of interaction becomes:
\begin{align}
E_I \;=\; - \frac{q^4}{16 \pi^3 r^2} \, \int_0^\infty \,d\nu \, 
\big| \widetilde{G}(\nu)\big|^2 \,e^{- 2 \nu r} \, 
\Big( \nu^4 + \frac{ 2 \nu^3}{r} + \frac{ 5 \nu^2}{r^2} + \frac{6 \nu}{r^3} 
+ \frac{3}{r^4} \Big) \;.
\end{align}

Introducing yet again the variable $u \equiv \nu r$,
\begin{align}
E_I \;=\; - \frac{q^4}{16 \pi^3 r^7} \, \int_0^\infty du \,e^{-2 u}
\, \big| \widetilde{G}(\frac{u}{r})\big|^2 \,  
\big( u^4 + 2 u^3 +  5 u^2 + 6 u + 3\big) \;.
\end{align}
When $\widetilde{G}(\nu)$ has a finite zero-frequency limit, we can extract
the long-distance behaviour of the interaction energy in terms of that
limit. Besides,
\begin{equation}
\widetilde{G}(0) \,=\,\frac{1}{3} \, 
\int_{-\infty}^{+\infty}\frac{d\nu}{2\pi} \, \langle
\big|\tilde{x}_i(\nu)\big|^2\rangle \;.
\end{equation}
Thus, the asymptotic form of the energy is:
\begin{align}
	E_I \;=\; - \frac{q^4}{16\pi^3 r^7} \, \big| \widetilde{G}(0)\big|^2 \, 
\int_0^\infty du \,e^{-2 u} \, 
\big( u^4 + 2 u^3 +  5 u^2 + 6 u + 3\big) \;.
\end{align}
By evaluating the integral, we get:
\begin{equation}
	E_I \;=\; - \frac{23}{(4 \pi )^3} \, \frac{q^4}{r^7} \,
	\big| \widetilde{G}(0)\big|^2.
\end{equation}
In the special case of the three-dimensional harmonic potential we just
considered before, we have, $\widetilde{G}(0) = \frac{1}{m \Omega^2}$, 
which reproduces our previous result.

Finally, note that we can also find the result for the London limit in the case
of a general central potential, since in that approximation we get:
\begin{align}\label{eq:l2}
	E_I \;=\; - \frac{3}{4} \,  \big(\frac{q^2}{4\pi}\big)^2 \frac{\pi}{r^6} \, 
	\int_0^\infty d\nu \, \big| \widetilde{G}(\nu)\big|^2 \;,
\end{align}
which again produces the right result for the harmonic potential case.

\section{Conclusions}\label{sec:conc}
We have presented a derivation of some known expressions for the
Van der Waals interaction energy between two atoms, based on a microscopic
description of the system, and applying functional methods: the energy is
obtained by functional integration of the degrees of freedom in the
Euclidean formalism, in order to obtain the vacuum energy from the
resulting effective action. 

We have analyzed the region where the description begins to fail, namely,
when the atoms are too close, and the dipole interaction may overcome the
binding energy of the electrons to their respective nuclei. 
This phenomenon shows up as the emergence of an imaginary part in the
energy, and the consequent vacuum decay probability per unit time.

We suggest that beyond such a limit a molecular description should be the
proper framework to describe the physics of the system.

Our results may be interpreted as providing a lower bound for the distances
to which one can apply the usual Van der Waals description, in terms of parameters
related to the structure of the atoms (in our case, $m$ and $\Omega$), and
the electromagnetic coupling ($q$). For distances larger that what we denoted by
$r_1$, an effective description for the interaction energy like the one we have
used should be reliable.


\section*{Acknowledgements}
The authors thank ANPCyT, CONICET and UNCuyo for financial support.

\end{document}